\documentclass[]{spie}  

\usepackage[utf8]{inputenc}

\usepackage{amsmath,amsfonts,amssymb}
\usepackage{graphicx}
\usepackage[colorlinks=true, allcolors=blue]{hyperref}
\usepackage{upgreek}

\title{Nanostructures for in-situ surface-enhanced Kretschmann-Raether ellipsometry}

\author[a,d]{D Mukherjee}
\author[a,b]{B Kalas}
\author[c]{S Burger}
\author[a]{G S\'afr\'an}
\author[a]{M Ser\'enyi}
\author[a,d]{M Fried}
\author[a,e]{P Petrik}
\affil[a]{Institute for Technical Physics and Materials Science, Centre for Energy Research, Konkoly-Thege Rd. 29-33, 1121 Budapest, Hungary}
\affil[b]{Doctoral School of Physics, Faculty of Science, University of P\'ecs, 7624 P\'ecs, Ifj\'us\'ag \'utja 6, P\'ecs, Hungary}
\affil[c]{Zuse Institute Berlin (ZIB) \& JCMwave GmbH, Takustrasse 7, 14195 Berlin, Germany}
\affil[d]{Doctoral School of Material Sciences and Technologies, \'Obuda University, N\'epsz\'inh\'az u. 8, 1081, Budapest, Hungary}
\affil[e] {Department of Electrical and Electronic Engineering, Institute of Physics, Faculty of Science and Technology,University of Debrecen, 4032 Debrecen, Hungary}

\authorinfo{Further author information: \\ (Send correspondence to email address: petrik.peter@ek-cer.hu) (Dr. P. Petrik).}

\begin{document} 

\maketitle

\begin{abstract}

Spectroscopic ellipsometry is a sensitive and optical model-supported quantitative tool to monitor interfaces. In this work, solid-liquid interfaces are studied using the Kretschmann-Raether configuration for biosensing applications. The interface layers support two purposes simultaneously: (i) chemical suitability for the adsorption of molecules to be detected and (ii) the optical enhancement of the signal to increase the sensitivity. Ellipsometry is not only used as a sensor but also as a quantitative measurement tool to study and understand the interface phenomena, and to develop the sensing layers for the largest possible optical sensitivity. Plasmonic and structured layers are of primary importance in terms of optical sensitivity. Layers structured both in lateral and vertical directions have been studied. Optical models based on both the transfer matrix and the finite element method were developed and used for the structural analysis where the material and geometrical derivatives are used in the inverse fitting process of the model data to the measurement. Structures utilizing plasmonic, diffraction, multilayer field enhancement, and other methods were analyzed as possible candidates for the improvement of the optical performance of the cell. Combinatorial and periodic plasmonic surface structures were developed to enhance the sensitivity of in-situ ellipsometry at solid-liquid interfaces utilizing the Kretschmann-Raether (KR) geometry. Ag$_x$Al$_{1-x}$ layers with variable compositions and Au layers with changing periods and critical dimensions were investigated to improve the performance of sensors based on the KR arrangement. 

\end{abstract}


\emph {This paper will be published in Proc. SPIE Vol. 12428 (2023) 124280S (Photonic Instrumentation Engineering X; DOI: 10.1117/12.2649080) and is made available as an electronic preprint with permission of SPIE. One print or electronic copy may be made for personal use only. Systematic or multiple reproduction, distribution to multiple locations via electronic or other means, duplication of any material in this paper for a fee or for commercial purposes, or modification of the content of the paper are prohibited.}

\section{INTRODUCTION}
\label{sec:intro}  

The research of nanostructures has been studied for many years, mostly due to their versatile and advantageous properties. Nanostructures have also opened up a realm of possibilities regarding scientific and medical research one of which is plasmonic nanostructures. In general, plasmonics is the field of research and technology that relies on the collective oscillation of free electrons in response to electromagnetic radiation in metallic thin films or various metallic nanostructures \cite{WIJAYA2011208}. Hence, plasmonic nanostructures are nanostructures capable of generating and controlling light at the nanoscale. These nanosized structures are capable of manipulating light waves in a wide range of directions due to their small size and shape. Surface plasmon polariton (SPP) resonance is observed at the interface of metal layers and dielectrics (e.g. air or liquids) due to the interaction of plasmons with light. This originates from a strongly confined surface wave that propagates along the interface and decays exponentially in both the metal film and the dielectric ambient.

The chemical composition of the plasmonic material is modified by alloying various metals for subtle customization of the optimum wavelength range for SPP \cite{Lithography,Noble,Structure,Band,Alloying}. It is an effective method for tuning the optical response of metallic thin films and nanostructures. Despite the well-known optical losses introduced by interband optical transitions \cite{Searching,PhysRevB,Peale08}, noble metals such as Ag, Au, and Pt have been preferred for SPP applications due to their abundant free electrons. Earth-abundant metals, such as Al and Mg, have recently been used as a new class of materials for low-optical-loss and low-cost optical components \cite{article10}. Metals such as Al and Cu are ideal for on-chip nanophotonics applications because of their fully complementary metal-oxide-semiconductor (CMOS) compatibility, combined with their abundance on Earth, results in low-cost device processing \cite{Alloying}. Furthermore, because of the high free electron densities found in metal alloys and intermetallics, they are promising candidates for alternative plasmonic materials \cite{Searching}.

Metal alloying via physical deposition methods results in nearly arbitrary tunability of their optical behavior in the ultraviolet-visible-near infrared (UV-Vis-NIR) wavelength range. This flexible spectrum control has also enabled the advancement of a wide range of advanced optoelectronic devices, including switches \cite{article11} and biosensors \cite{article12}. It has previously been demonstrated that combining Ag or Au with another metal that contributes two or three electrons per atom to the free electron gas can significantly alter the reflection and absorption spectra. The investigation of single-phase Ag-Cu alloys with various compositions led to the conclusion that the optical properties of these alloys could be efficiently manipulated by the annealing temperature and composition, making Ag-Cu an appealing material for plasmonics \cite{Yang13}. Optical spectroscopy has been used to investigate the dielectric properties of a variety of alloy systems, including Au-Ag \cite{Noble,Pena}, Ag-Cu \cite{Noble,Yang13}, and Au-Cu \cite{Noble}, as well as a combinatorial gold-aluminium (Ag$_x$Al$_{1-x}$) layer \cite{Combinatorial}.

Of all the materials in the Vis-NIR region, Ag possesses the best characteristics for plasmonic applications \cite{Blaber_2010} and is widely accessible. Ag, as is well known, develops according to the Volmer-Weber growth model \cite{Sennett50}, in which the deposited Ag atoms first form solitary islands. These islands expand further as the deposition goes on, eventually joining to form a semi continuous layer. A thin Ge layer of 1-2 nm can be placed underneath the Ag layer to solve this issue \cite{Ultrasmooth,Zhang}. Using this technique, the percolation threshold and surface roughness value for Ag nanofilms have both decreased \cite{FAHLAND2001334}. However, Ge exhibits strong visible wavelength absorption, which consequently reduces transmittance. This issue has been addressed by the deposition of an extremely thin and smooth Ag sheet, which exhibits low optical loss and low electrical resistance \cite{Cheng}. This film was created by co-depositing a very small amount of Al during the deposition of Ag without the use of a wetting layer. A modest amount of Al was shown to restrict the growth of Ag's 3D islands in the same investigation, leading to the production of an ultra-thin layer with a reduced surface roughness of 1 nm and a reduced percolation threshold of 6 nm. Ion-implanted Al-Ag bimetallic substrate has been recommended as a suitable instrument for surface enhanced fluorescence (SEF) detection as a potential application \cite{article26}.

In optical reflector applications, an Al layer with a high reflectivity across the Vis spectral range has been used \cite{palik1998handbook}. A 7-nm Al-doped Ag film's ability to withstand ambient conditions for more than six months without a protective layer has also been demonstrated \cite{High}. When compared to pure Ag, Al-doped Ag has more advantages, including the capacity to create ultrathin films, improved thermal and long-term stability, better substrate adherence, and improved 3D nanostructure coverage \cite{High,article31}. Additionally, a straightforward annealing procedure can minimize its optical loss even more. It has also been demonstrated that altering annealing temperatures and Ag concentrations in Al-Ag alloys can change the interband transition's center position, which is around 1.5 eV \cite{Dielectric,YANG201323}. All of these advantages make it easier to create long-range surface plasmon polariton (LR-SPP) waveguides \cite{High} and high-performance Al plasmonic devices \cite{Diest13,article35}.

Typically, a thin Au film is utilized as the sensing layer in SPR spectroscopy. At the intersection of the Au layer and aqueous environment, incoming light can be used to excite propagating surface plasmon oscillation using the Kretschmann-Raether arrangement \cite{KretschmannRaether+1968+615+617}. A dip in the reflectance spectrum results from the incident light coupling with surface plasmons under the right circumstances. The thickness ($d$) and optical characteristics of the Au layer \cite{KALAS2017585}, the angle of incidence ($\theta$) of the light beam, the optical characteristics of the configuration, and -- most crucially -- the optical characteristics of the investigated ambient close to the Au surface -- all have a significant impact on the precise wavelength ($\lambda$) value of this dip. The increased sensitivity can also be facilitated by the creation of unique layer architectures \cite{PhysRevB78}. A better sensing performance may be achieved by utilizing not only a bare Au layer but additionally one or more 2D layers (such as graphene or molybdenum-disulfide) on top of the Au film \cite{C6CS00195E}. The so-called long-range surface plasmons are used in another layer structure with increased sensitivity (LRSPRs). Similar to SPR, electromagnetic waves (also known as Bloch surface waves) are limited to the layer structure's surface and exhibit an exponential field decay both inside the layered medium and in the surrounding liquid. In comparison to the often utilized Au layers, these customized periodic layer architectures have a number of enhanced features.

The Kretschmann-Raether configuration is based on ellipsometry principles where light is reflected at an interface between a solid and a liquid, allowing for the study of the optical properties of materials. This configuration consists of a sample placed between two prisms that are used to measure the angle-dependent reflectance and transmittance, which is then used to calculate the absorption and refraction of the sample. The core is usually made of a material with a higher refractive index than the cladding material, which enables light to be confined in the core. It relies on the use of a polarized light source, which creates elliptically polarized light as it passes through the waveguide. This light can then be detected after it has passed through the waveguide and its properties can be used to analyze and measure the properties of the waveguide itself. It also has the advantage of providing good transmission with small bending losses, as well as being easy to fabricate. 

Nanostructures have been successfully employed for the purpose of in-situ surface-enhanced Kretschmann-Raether ellipsometry. This type of optofluidics technology allows for the non-invasive transduction of evanescent fields to surface plasmon polariton in order to carry out ellipsometric measurements. In recent years, these nanostructures have been used to improve the resolution of this type of spectroscopy and to reduce the time required for data analysis. This is a significant advancement, as it enables researchers to to measure both optical and topographical properties of a sample and study surface interactions in real-time, without having to remove the sample from the measurement environment. This approach is also capable of providing rapid and accurate information on the surface morphology, structure, and composition of the sample.

We introduce a scaled-up dual DC magnetron sputtering apparatus that deposits a thin and smooth combinatorial Ag$_x$Al$_{1-x}$ alloy sheet with x covering the entire composition range 0 $<$x$<$ 1 \cite{article36}. Variable angle spectroscopic ellipsometry (VASE) was used to analyze the optical characteristics of the deposited Ag$_x$Al$_{1-x}$ alloy sheet over a broad wavelength ($\lambda$) range. On a fused silica (FS) substrate, the intermetallic layer was created, and it was then covered with an RF-sputtered Si$_3$N$_4$ layer. In addition to operating as a waveguide layer to create a coupled plasmon-waveguide resonator (CPWR) structure in the Kretschmann-Raether (KR) configuration \cite{Long,NENNINGER2001145,Vernoux18,JING2019103}, this extra layer also protects the metal layer/air interface from contact \cite{KretschmannRaether+1968+615+617}.

\section{MATERIALS AND METHODS}

A dual-source rotating compensator spectroscopic ellipsometer named as  J.A. Woollam M-2000DI was used in the wavelength range of $\lambda$ = 190-1700 nm, variable utilizing the Kretschmann-Raether (KR) geometry for enabling $\theta$ up to 75$^{\circ}$ for focus extension. An improved hemisphere has been used for the KR ellipsometry (KRSE) setup for providing an optimal signal-to noise ratio in the crucial spectral range below 300 nm.

The mapping stage is required for the optical adjustment of the KR setup which enables to use the optical parameters of  the KR setup (focusing lenses, hemisphere, glass slide, index matching liquid) at a spectral range of 190–1700 nm of the ellipsometer. The theoretical spectral resolution bandwidth is expected to be around 5 nm and 10 nm in the UV/VIS and in the near infrared wavelength ranges, respectively. The experimental spectral density of the experimental data points is found to be around 1.6 nm and 3.4 nm in the UV/VIS and in the near infrared wavelength ranges.

\begin{figure} [ht]
   \begin{center}
   \begin{tabular}{c} 
   \includegraphics[height=5cm]{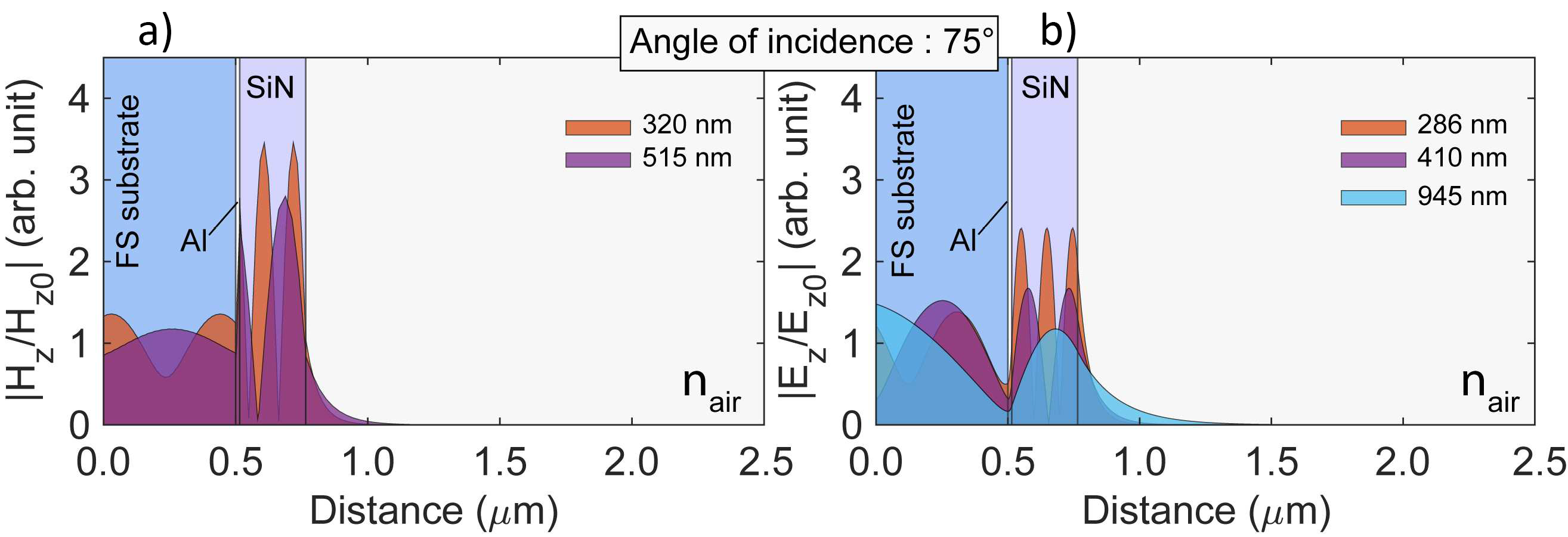}
   \end{tabular}
   \end{center}
   \caption{\label{FEM-fields}Near-field distribution calculated for p- (a) and s-polarizations (b) using the finite element method for the Ag$_x$Al$_{1-x}$ structure at $x=0$, with thicnesses of the plasmonic and Si$_3$N$_4$ waveguide layers of 15 and 250 nm, respectively. E$_z$ and H$_z$ denote the z-components of the electric and magnetic fields, respectively. The areas with different colors correspond to different wavelengths.}
   \end{figure}

Levenberg–Marquardt algorithm\cite{osti_7256021} is used to analyze the optical properties and other physical parameters by fitting the parameters of an optical model by utilizing the transfer matrix method. To capitalize on the potential of SE for in situ TIRE measurements, a 10-L flow cell with a KR configuration consisting of an FS hemicylinder was used. This configuration allows for the investigation of optical properties in a liquid environment with a wavelength range of $\lambda$ = 190-1700 nm, as well as a wide angle of incidence range of $\theta$ = 45-75$^\circ$. A focused light beam with a spot size less than 1 mm is used during the measurements to ensure the best performance (e.g., to ensure a normal incidence at the air/hemisphere interface over the entire illuminated spot). Although the beam cannot be focused below a diameter of approximately 300 microns with the current hardware, if the scanning capability is not used, the lateral size of the flow cell can theoretically be as small as the spot itself, with a depth also less than a millimeter, resulting in a microliter-size cell. The spot size can also be reduced when the wavelength range is limited. This method can also be combined with imaging ellipsometry, which has a lateral resolution of one micron. The distribuion of near-field was calculated using the finite element method (Fig. \ref{FEM-fields}).

Combinatorial film was deposited by "single-sample" micro combinatory at room temperature on a 25 mm x 10 mm (width x length) and $150 \pm 25$ $\upmu$m thin UV-grade FS substrate (purchased from Valley Design Corp). The deposition produced a layer with a gradient composition of Ag$_x$Al$_{1-x}$ ranging from 0 $<x<$ 1. A 20 mm long gradient track is enclosed between 2.5 mm long lead-in sections of one target's flux on the 25 mm long substrate. The current arrangement sweeps a shutter with a 1 mm x 10 mm slot above the substrate in fine steps, while the power of the two-magnetron sources is regulated in sync with the slot movement. The fluence of Ag gradually decreases from 100 to 0 as the slot passes over the substrate, whereas that of Al gradually increases from 0 to 100, resulting in the necessary gradient of the composition. For the Ag and Al targets, the maximum applied power values were 150 W and 330 W, respectively.

In Fig. \ref{FEM-fields}, the 15-nm Al/250-nm Si$_3$N$_4$ structure and the FEM-calculated EM fields are presented at the discrete resonance wavelengths based on the arrangement. From this analysis it is apparent that the maximum values of the normalized E$_z$/E$_{z0}$ and H$_z$/H$_{z0}$ fields are usually close to the Si$_3$N$_4$/air interface and thus they can be used for optical sensing applications. The various decay length of these fields can be useful during optical detection since the separation of processes near to the surface (e.g., molecule adsorption) are easier from the changes in the bulk material.

In case of Al resonances, wavelength positions are available from 265 nm to 1504 nm, thus covering almost the entire simulated   spectra. For both Al and Ag, the full width at half maximum (FWHM) values can be adjusted well below 10 nm. For the simulations being described, the dielectric functions $\epsilon$ = $\epsilon_1$ - i$\epsilon_2$ [= N${^2}$ = (n - ik)${^2}$] of Ag and Al are imported \cite{LYNCH1998341}. The optical properties of Si$_3$N$_4$ derived from the ellipsometric measurement of a single Si$_3$N$_4$ layer deposited on Si substrate. The refractive index of air and glass were chosen to be       n$_{air}$ = 1 and n$_{glass}$ = 1.5, respectively. Here $\epsilon_1$, $\epsilon_2$ and N denote the real- and the imaginary parts of the complex dielectric function and the complex refractive index, respectively.

\begin{figure} [ht]
   \begin{center}
   \begin{tabular}{c} 
   \includegraphics[height=5cm]{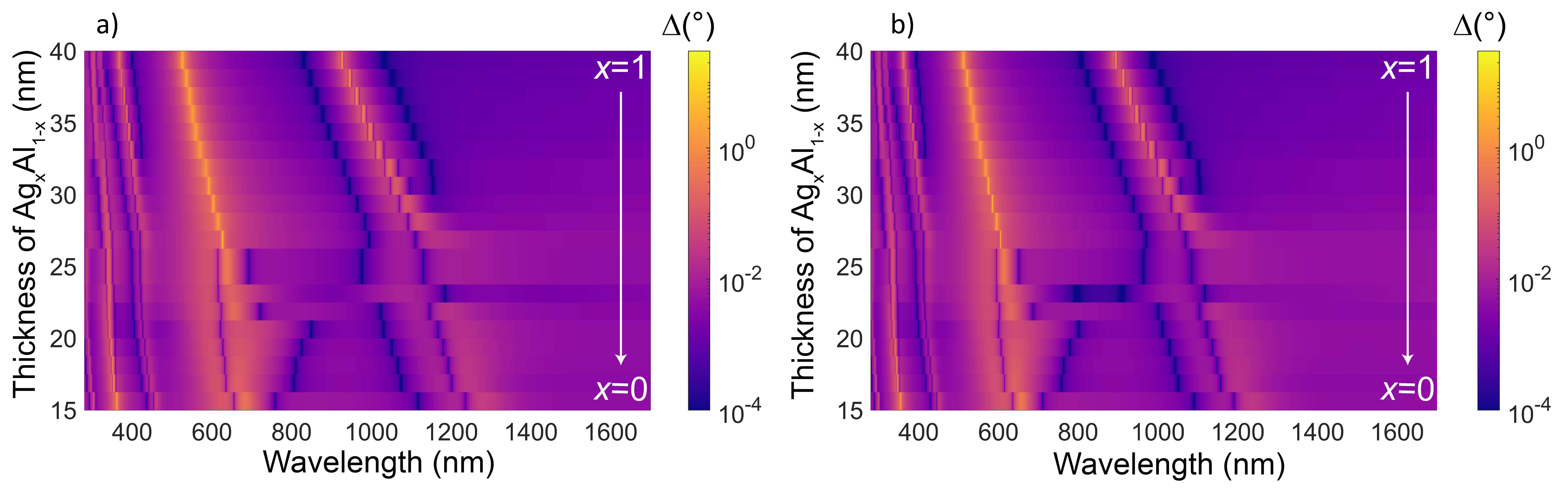}
    \end{tabular}
    \end{center}
   \caption[example] 
   { \label{Delta-vs-d} 
Variation of the ellipsometric phase shift ($\Delta$ = arg($r_p$/$r_s$)) as a funtion of the thickness of the Ag$_x$Al$_{1-x}$ layer and the wavelength for angles of incidence of 71$^\circ$ (a) and 73$^\circ$ (b).}
   \end{figure}

The precise regulation of the slot movement helped to achieve the variable thickness as well, which can be utilized to adjust the wavelength of highest sensitivity (Fig. \ref{Delta-vs-d}). The following are the key benefits of the combinatorial deposition technique: (i) All sample preparation parameters and substrate properties, with the exception of the modulated parameter (the composition and/or thickness), are guaranteed to be the same when the deposited layer is prepared in a single process step. (ii) Because a lateral scan can be used to automatically execute the optical measurement and evaluation, they are also simpler and more effective. A single-process interpretation is supported by the evaluation procedure as well. Without making any lateral dependence assumptions, the combinatorial technique also enables the characterization of more significant modulations and unexpected variations of the attributes.

\section{RESULTS AND DISCUSSION}

In the KR-cell filled with air, SE analysis of the Ag$_x$Al$_{1-x}$/Si$_3$N$_4$ structure was performed parallel to the long edge of the sample with a lateral resolution of 1 mm. Considerable spectrum changes are observed in the measured spectra at $\theta$ = 47$^\circ$ across the wavelength range of 190-1700 nm, showing a significant impact of the varying metal composition. For optical sensing setups where the AOI cannot be modified, the significance of adjusting the spectral position of the resonance peaks by merely changing the light spot position along the sample can be advantageous. For pure Ag, it is also true that resonance peak positions have a lower limit regarding the smallest wavelength value that can be accessed, and that this restriction can only be further decreased by changing its optical characteristics. Additionally, the broadening typically degrades more quickly when the AOI changes than when the composition does. There are also wavelength ranges that cannot be accessed by adjusting the AOI. From the fitted spectra that were obtained from measurements made in the KR-cell, the composition-dependent dielectric functions were determined. The thicknesses of the Ag$_x$Al$_{1-x}$/Si$_3$N$_4$ layers and the oscillator parameter values were fitted in this investigation.

It should be observed that when $x$ decreases, $\epsilon_2$ increases. This can be explained by the rise in Al impurities (more electron scattering as a result of the system's compositional instability) \cite{Auer, DeSilva_2017}. Another well-known consequence \cite{DeSilva_2017} is the emergence of a new $\epsilon_2$ peak near 500 nm, as well as the blue-shifting of the interband transitions caused by decreasing $x$. New interband transitions are responsible for the latter phenomena. Once more, a considerable change in $\epsilon_2$ can be seen starting from the Al-rich side (x $\approx$ 0). Here also, the $\epsilon_2$ -peak swings toward shorter wavelengths, and it is clear that in the NIR wavelength range, the value of $\epsilon_2$ decreases as the value of $x$ increases. In actuality, the peak of the interband transition moves from 740 nm (x $\approx$ 0) to 485 nm (x $\approx$ 0.35). It is crucial to note for the clarification of these results that variations in the content and thickness of the combinatorial layer have a significant impact on the optical properties \cite{PRIBIL2004443,article67,TODOROV201722}. It is widely known that metal film thickness, particularly when it is less than about 50 nm \cite{Drachev08}, can greatly affect the optical properties in addition to the crystallites/grain structure and size. This effect can be related to the fact that the average route of free electrons in Ag has a typical value of l = 44 nm \cite{cai2010optical}, which is comparable to the crystallite size. The average free path of the electrons is thus influenced by the grain size, and as a result, the values and so-called free electron effect will be visible \cite{UKreibig_1974}. In addition, it has been demonstrated in the past that the conditions of Al layer deposition using different procedures can have a significant impact on the dielectric function of the film because they can change the usual grain size over a large range (about 10–50 nm)\cite{VanGils,SUN20076962,Zhou_2015}. Among the observed points, layers with compositions $x$ = 0.40, 0.45, and 0.50 show the most distinctive properties, according to the measured psi-spectra and the findings from the optical analysis. Dielectric functions in this $x$-range were either described by a single Drude-oscillator (x $\approx$ 0.4) or by a Drude-Gauss method (x $\approx$ 0.45, 0.50). Phase alterations may be attributed for this phenomena in the aforementioned composition range. \cite{Mao}

The compositions $x = 0$ (pure Al) and $x = 1$ (pure Ag) of the produced sensor structure had optimum thicknesses because the optical characteristics of these pure phases were thought to be more dependable than the references for mixed phases. Although it would be possible to tune the lateral thickness profile after determining the optical characteristics for all values, this one-sample combinatorial device is now only optimized for linearly graded compositions and thicknesses. TMM calculations were conducted for investigating the change in $\Delta$ ellipsometric angle for $\Delta$n$_{water}$= 10$^{-4}$. From these figures it is evident that the higher sensitivity can be achieved at the lower AOI value (71$^\circ$) since this AOI is the closest to the corresponding value of the total internal reflection angle. The highest sensitivity can be achieved by using $x = 1$ composition and its value significantly decreases by increasing the amount of Al. This effect can be explained as the higher absorption of these compositions, thus requiring further thickness optimization in order to enhance the sensing performance in this $x$-range.

\begin{figure} [ht]
   \begin{center}
   \begin{tabular}{c} 
   \includegraphics[height=5cm]{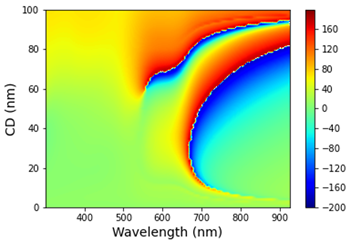}
    \end{tabular}
    \end{center}
   \caption{\label{Delta-vs-CD} $\Delta$ ellipsometric angle calculated for different critical dimension (CD) and wavelength values in the KR configuration for a gold layer on glass with a thickness and period of 200 nm and 100 nm, respectively.}
   \end{figure}

It was shown that the phase information determined by ellipsometry together with the help of variable resonant features can be utilized in both combinatorial Ag$_x$Al$_{1-x}$ and periodic Au grating structures (Fig. \ref{Delta-vs-CD}) prepared by sputtering on glass plates that were attached to a hemi-cylinder in the KR geometry. The phase is measured as $\Delta$ = arg($r_s$/$r_p$), where $r_p$ and $r_s$ are the reflection coefficients of light polarized parallel and perpendicular to the plane of incidence, respectively. Using a Woollam M-2000DI ellipsometer the resolution in space (focused spot) and time is typically 0.3 mm by 0.9 mm and 1 s, respectively, while the wavelength range is $\lambda$ = 190-1700 nm. At an optimum plasmonic configuration, sensitivities of $10^{-6}$ and 10 pg/mm$^2$ can be achieved in refractive index and surface mass density units, respectively. To develop phase retrieval and sample parameter reconstruction method, Au gratings were created and modeled using finite element (JCMwave) and transfer matrix methods. They are of primary importance for the identification of grating parameters that are most promising in experimental investigations.

\section{CONCLUSION}

Here, a combinatorial thin layer structure that can optically detect minute changes in the refractive index of the surrounding gas or liquid was investigated. The Ag$_x$Al$_{1-x}$/Si$_3$N$_4$ layer arrangement that makes up the combinatorial sample was deposited on an FS glass slide using dual DC magnetron sputtering. It was demonstrated that a single sample may be used to tailor the thickness of the Ag$_x$Al$_{1-x}$ and Si$_3$N$_4$ layers as well as the composition over the entire range of ($0 < x < 1$). KR SE was used to characterize the optical characteristics of the combinatorial layer, from which the composition-dependent trends in both $\epsilon_1$ and $\epsilon_2$ were found, where the changes are in good agreement with previously published research. The structure's sensing capabilities for gas sensing and biosensing in a liquid environment were also examined numerically (at composition values of $x = 0$ and $x = 1$). The structure's ability to detect highly sensitively in the UV, Vis, and NIR wavelength ranges at a single AOI was confirmed based on the findings. Ellipsometric phase-sensitive measurements with detection limit values of 4.1$\times$10$^6$ RIU and 7.9$\times$10$^6$ RIU at the Ag- and Al-sides of the chip were also performed. Periodic Au grating structures were created and modeled to develop phase retrieval and sample parameter reconstruction method which is extremly beneficial for the identification of grating parameters that are most promising in experimental investigations.

\section{ACKNOWLEDGEMENTS}

Support from the National Development Agency Grants of OTKA Nr. K131515 is greatly acknowledged. The work in frame of the 20FUN02 ‘‘POLight’’ project has received funding from the EMPIR programme co-financed by the Participating States and from the European Union’s Horizon 2020 research and innovation programme. Project no. TKP2021-EGA04 has been implemented with the support provided by the Ministry of Innovation and Technology of Hungary from the National Research, Development and Innovation Fund, financed under the TKP2021 funding scheme.

\bibliography{nanoKRSE} 
\bibliographystyle{spiebib} 

\end{document}